\documentclass[apj]{emulateapj}
\usepackage{apjfonts}
\usepackage{amsbsy}

\newcommand{\kms}{{\rm km\, s}^{-1}}
\newcommand{\kpc}{\ {\rm kpc}}
\newcommand{\up}[1]{{\rm #1}}
\newcommand{\SSS}{\S}
\newcommand{\bdv}[1]{\pmb{#1}}

\begin{document}

\title{Halo Structures of Gravitational Lens Galaxies \altaffilmark{*}}

\author{Jaiyul Yoo\altaffilmark{1}, Christopher S. Kochanek\altaffilmark{1}, 
Emilio E. Falco\altaffilmark{2}, and Brian A. McLeod\altaffilmark{2}}

\altaffiltext{*}{Based on Observations made with the NASA/ESA 
{\it Hubble Space Telescope}, obtained at the Space Telescope Science 
Institute, which is operated by AURA, Inc., under NASA contract NAS5-26555.}

\altaffiltext{1}{Department of Astronomy, The Ohio State University, 
140 West 18th Avenue, Columbus, OH 43210;
jaiyul@astronomy.ohio-state.edu, ckochanek@astronomy.ohio-state.edu}

\altaffiltext{2}{Harvard-Smithsonian Center for Astrophysics, 60 Garden 
Street, Cambridge, MA 02138; efalco@cfa.harvard.edu, bmcleod@cfa.harvard.edu}

\slugcomment{accepted for publication in The Astrophysical Journal}
\shorttitle{HALO STRUCTURES OF GRAVITATIONAL LENS GALAXIES}
\shortauthors{YOO ET AL.}

\begin{abstract}
We explore the halo structure of four gravitational lenses with 
well-observed, thin Einstein rings. We find that
the gravitational potentials are well described by 
ellipsoidal density distributions in the sense that the 
best-fit nonellipsoidal models have parameters consistent with their
ellipsoidal counterparts. We find upper limits on the standard parameters for
the deviation from an ellipse of
$|a_3^\up{B}/a_0|<0.023$, 0.019, 0.037, and 0.035, and
$|a_4^\up{B}/a_0|<0.034$, 0.041, 0.051, and 0.064
for SDSS~J0924+0219, HE~0435$-$1223, B~1938+666, and PG~1115+080,
respectively. We find that the lens galaxies are at the centers
of their dark matter halos, and obtain upper limits for the offset of 
each center of mass from the center of light of 
$|\Delta\bdv{x}|<0\farcs004$, $0\farcs005$, $0\farcs009$, and $0\farcs005$, 
corresponding to 22, 29, 70, and 23~pc. These limits also exclude
the possibility of any significant lopsidedness of the dark matter halos and
set an upper limit of $f_\up{sat}\lesssim\sqrt{N}\%$ on the mass fraction of 
massive substructures inside the Einstein ring if they are divided over
$N$ satellites.
We also explore the properties of galaxies as substructures in groups for the
lens PG~1115+080, finding evidence for dark matter halos associated with the
galaxies but no evidence for a clear distinction between satellite and 
central galaxies.
\end{abstract}


\section{Introduction}
\label{sec:int}
In cold dark matter (CDM) models, a typical virialized object consists of
a luminous galaxy centered in a dark matter halo and surrounded by satellites,
some of which may contain no stars (e.g., \citealt{james,tumul}).
In this paper, we explore three questions
about the structure of such halos
using four gravitational lenses in which we can
accurately measure the structure of the Einstein ring image of the quasar host
galaxy.

First, we explore the angular structure of the dark matter halos.
The angular structure of
luminous early-type galaxies is well-approximated by projected ellipsoidal
distributions, or ellipses, albeit with small systematic deviations 
\citep{iso2,rest}.
Considerably less is known about the angular structure of dark matter halos
(e.g., \citealt{neal,shape}),
particularly after their structure is modified by the dissipative cooling 
and reheating of the baryons by star formation as the galaxy evolves.
This issue is relevant to the halo structures less from an interest in the 
ellipticity of the halos than
from the possibility that strong deviations from an ellipsoidal density
distribution may be responsible for the flux ratio anomalies in
gravitational lenses (e.g., \citealt{mao,zhao,chi,benton})
that otherwise provide the strongest 
evidence for the existence of dark matter substructures in the
halos of galaxies \citep{neal1}.
\citet{wyn2} demonstrated that lens potentials
with arbitrary angular structure could reproduce the 
flux anomalies without the substructures, but \citet{chris3}, \citet{ring}, and
\citet{congdon} demonstrated for several lenses that the deviations 
in the lens potentials
from an ellipsoidal distribution are too small to remove the flux
anomalies.  Here we extend our study of the shape of the gravitational 
fields in galaxy halos to three more gravitational lens systems.

Second, we explore whether the center of mass of the luminous galaxy is 
identical to the center of mass of the halo, since most
popular halo models assume that luminous galaxies are
centered on their dark matter halos (e.g., \citealt{uros,dhw,cooray}). 
Although it is unlikely that the 
core of the stellar and dark matter distributions can be offset in a galaxy, 
potentially there could be some lopsidedness or other effects that would make 
it invalid to assume that the center of light is the center of mass.
Weakening the centering of the halo on the lens galaxy was also one of the
factors permitting \citet{wyn2} to explain flux ratio anomalies with 
changes in the lens potential.
Small numbers of massive, but dark substructures would also create an offset
between the centers of mass and light.

Finally, we explore the halo structure of a galaxy group. 
It is likely that central and satellite galaxies in a group
show significantly different mass-to-light ratios on large scales due to the
mass differences between 
halos associated with each galaxy (e.g., \citealt{zz}).
This could explain, for example, the increasing evidence that
early-type galaxies are heterogeneous in their structure (e.g.,
\citealt{chrisnew}).
Most lens galaxies are members of small galaxy groups, where the
other galaxies in the group can perturb the lens through their
tidal gravities. If we consider a lens centered on one of the 
satellite galaxies, then we might plausibly identify the central
galaxy if one of the other galaxies in the group has a
significantly higher mass-to-light ratio on large scales
than the others.  If, on the other hand, the structural
differences between central and satellite galaxies are
small compared to the intrinsic scatter, then no such 
identification can be made. We can explore these possibilities using
lens galaxies in small groups.

We examine these three issues 
using four gravitational lenses with well-defined Einstein ring images of
the quasar host galaxy:
SDSS~J0924+0219 \citep{inada,chucksdss},
HE~0435$-$1223 \citep{wis,nick,chrisnew}, B~1938+666 \citep{king,king2,ton}
and PG~1115+080 \citep{wey,courbin,keeton,impey,treu}.
We use the shape of the Einstein ring formed by quasar host galaxies 
to accurately constrain
the angular structure of the underlying potential \citep{chris2,ring}.
In \SSS\ref{sec:obs} we briefly present the {\it Hubble Space Telescope}
($HST$) observations of the four lenses followed by a
short summary of the lens models in \SSS\ref{sec:model}. Our main results
on the first two questions are presented in \SSS\ref{sec:result1}. We discuss
the third question for the galaxy group surrounding PG~1115+080 in
\SSS\ref{sec:result2}. We summarize our results in \SSS\ref{sec:sum}.

\section{Observations}
\label{sec:obs}
We analyzed four gravitational lenses (SDSS~J0924+0219, 
HE~0435$-$1223, B~1938+666, and
PG~1115+080) with well-defined, thin Einstein ring images of the quasar host
observed at $H$-band (NIC2/F160W) with the Near-Infrared Camera and 
Multi-Object Spectrograph on the $HST$. The data were
reduced and modeled using our standard procedures \citep{brian,lehar}, and
we extracted the Einstein ring curve as described in 
\citet{chris2}. The image is fitted by a model consisting of the lens galaxy,
the quasar images, and a lensed host galaxy. The lens galaxy and quasar images
are then subtracted, and we measure the radius $r(\theta)$ of the Einstein ring
as a function of position angle around the lens galaxy. The radius $r(\theta)$
of the ring curve can then be used as a constraint on lens models under the
assumption that the isophotes of the host galaxy are close to ellipsoidal.

SDSS~J0924+0219 consists of four lensed quasar images of a source at 
$z_s=1.524$ produced by an elliptical galaxy at $z_l=0.393$ 
\citep{ofek,cosmo} and
the lens is well-separated from nearby galaxies \citep{inada,chucksdss}. 
Eight dithered
images of the Einstein ring were obtained on 2004 June 1 for a total 
integration time of 4640 seconds (Morgan et al. in preparation).
We use the image positions from \citet{chucksdss}.
The lens is quite isolated and we use only an external shear to model its 
environment. By isolated, we mean that no nearby object provides higher-order
perturbations than an external shear.

The lens system HE~0435$-$1223 is composed of a four-image lensed quasar at
$z_s=1.689$ \citep{wis} and an elliptical lens galaxy at $z_l=0.4546$ 
\citep{nick,ofek}. A spiral-rich group of about 10 galaxies was found within 
40\arcsec~and the closest spiral galaxy G22 (SBb) is separated 
from the lens galaxy by 4\farcs46.
We use the positions of the lensed images 
and field galaxies around the lens from the $HST$/ACS images described in 
\citet{nick}, and the 2560s $HST$/NICMOS image described in 
\citet{chrisnew}. As in \citet{chrisnew}, we 
model the environment with an
external shear and the nearby galaxy G22.

B~1938+666 consists of a normal early-type lens galaxy of redshift $z_l=0.88$ 
\citep{ton} that produces a two-image system and a four-image system of a
radio source of unknown redshift \citep{king}. 
The host galaxy is seen as an almost perfectly circular ring around the lens
\citep{king2}. We analyze the 2816s NIC2 image taken on 1997 October 7.
For the radio image positions, we use the 5~GHz observations of \citet{king}.
We allow the registration of radio and optical coordinates to be optimized
as part of the fit.
The lens is fairly isolated and we model the environment using only an external
shear.

PG~1115+080 consists
of four images of a $z_s=1.72$ quasar lensed by a $z_l=0.31$ elliptical galaxy
\citep{wey,impey}.
This system belongs to a small group of galaxies producing non-negligible 
higher-order perturbations. In \citet{ring}, like most previous studies, 
we modeled the
environment as a separate group halo. Here we explore models with halos
centered on the galaxy positions provided by \citet{impey}.
We use the image positions from \citet{impey}.

We use a concordance cosmological
model with density parameters $\Omega_m=0.3$, $\Omega_\Lambda=0.7$
and the Hubble constant $h\equiv H_0/100~{\rm kms^{-1}Mpc^{-1}}=0.7$ 
to calculate angular diameter distances in a $\Lambda$CDM 
universe.

\begin{deluxetable*}{lcccccccc}
\tabletypesize{\scriptsize}
\tablewidth{0pt}
\tablecaption{Best-Fit Models}
\tablehead{\colhead{} & \multicolumn{2}{c}{SDSS~J0924+0219} & \colhead{} &
\multicolumn{2}{c}{HE~0435$-$1223} & \colhead{} & 
\multicolumn{2}{c}{B~1938+666} \\ 
\cline{2-3} \cline{5-6} \cline{8-9}\\
\colhead{Parameter} & \colhead{Ellipsoidal} & \colhead{Nonellipsoidal}
& \colhead{} & \colhead{Ellipsoidal} & \colhead{Nonellipsoidal}
& \colhead{} & \colhead{Ellipsoidal} & \colhead{Nonellipsoidal} }
\startdata
$q_s$............. & $0.86\pm0.06$ & $0.84\pm0.16$ & & $0.47\pm0.02$ 
                     & $0.51\pm0.06$ & &$0.56\pm0.10$ & $0.56\pm0.18$ \\
$\theta_s$(degs)... & $-24.4\pm14.3~~$ & $-28.0\pm22.7~~$ & 
                       & $-3.9\pm2.0$~~~& $-6.1\pm3.2$~~ & & $-71.2\pm9.3$~~~~ 
                       & $-71.2\pm23.3$~~~ \\
$\gamma$.............. & 
                       $0.058\pm0.013$ & $0.058\pm0.023$ & & $0.035\pm0.003$ &
                       $0.065\pm0.003$ & & $0.026\pm0.004$ & $0.026\pm0.007$ \\
$\theta_\gamma$(degs).. & 
           $65.6\pm0.9~$ & $62.0\pm3.5~$ & & $-60.2\pm5.1$~~~~ & 
           $-68.3\pm4.1$~~~~ & & $33.1\pm14.0$ & $33.3\pm15.6$\\
$\gamma_g$............ & 
             $-$ & $-$  & & $0.040\pm0.001$ & $0.042\pm0.003$ & & $-$ & $-$\\
$r_g$(arcsec)
      & $-$ & $-$ & & $\equiv4.46$ & $\equiv 4.46$ & & $-$ & $-$\\
$\theta_g$(degs).. & $-$ & $-$ & & ~~~~$\equiv-144.7$ & ~~~~$\equiv-144.7$ 
                      & & $-$ & $-$ \\
$x_L$(arcsec) & 
                $-0.001\pm0.001~~$ & $-0.001\pm0.003~~$ & & $-0.003\pm0.002~~~$
              & $-0.001\pm0.003$~~ & & $-0.007\pm0.007$~~ & 
                $-0.007\pm0.007$~~ \\
$y_L$(arcsec) & 
                $0.001\pm0.001$ & $-0.004\pm0.003~~$ & & $0.000\pm0.001$ & 
                $0.001\pm0.003$ & & $0.004\pm0.007$ & $0.004\pm0.009$\\
$q_L$............ & 
        $0.87\pm0.01$ & $0.85\pm0.02$ & & $0.62\pm0.01$ & $0.55\pm0.01$ & & 
        $0.89\pm0.02$ & $0.89\pm0.01$\\
$\theta_L$(degs).. & 
                      $-68.1\pm7.3~~~~$ & $-67.2\pm9.8~~~$ && $-14.3\pm9.1~~~~$
                    & $-10.5\pm9.4~~~~$ & & $-54.9\pm8.5$~~~~ &
                      $-54.5\pm10.2$~~~\\
$a_0$............ & 
       $0.88\pm0.01$ & $0.88\pm0.02$ & & $1.21\pm0.01$ & $1.21\pm0.01$ & 
      & $0.46\pm0.01$ & $0.46\pm0.01$\\
$10^2a_2$....... & 
            $-1.4\pm0.2~~$ & $-1.6\pm0.4~~$ & & $8.3\pm0.4$ & $11.1\pm0.5~$
          & & $-0.3\pm0.3$~~ & $-0.3\pm0.4$~~~\\
$10^2b_2$....... & 
            $-1.4\pm0.4~~$ & $-1.7\pm1.1~~$ & & $-4.5\pm0.4~~~$ & 
            $-4.2\pm0.7$~~ & & $-0.8\pm0.1$~~ & $-0.8\pm0.3$~~~\\
$10^3\Delta a_3$.... 
                 & $\equiv0$ & $-0.9\pm2.5~~$ & & $\equiv0$ & $2.1\pm3.3$
                 & & $\equiv0$ & $0.0\pm4.8$\\
$10^3\Delta b_3$.... 
                 & $\equiv0$ & $-2.2\pm3.8~~$ & & $\equiv0$ & $1.4\pm4.7$
                 & & $\equiv0$ & $0.1\pm4.3$\\
$10^3\Delta a_4$.... 
                 & $\equiv0$ & $0.8\pm3.0$ & & $\equiv0$ & $2.9\pm3.3$
                 & & $\equiv0$ & $0.0\pm2.1$\\
$10^3\Delta b_4$.... 
                 & $\equiv0$ & $1.2\pm3.3$ & & $\equiv0$ & $0.1\pm6.6$
                 & & $\equiv0$ & $0.0\pm1.7$\\
$10^4\Delta a_5$.... 
                 & $\equiv0$ & $-1.7\pm11.9~$ & & $\equiv0$ & $0.1\pm8.4$
                 & & $\equiv0$ & $0.1\pm1.2$\\
$10^4\Delta b_5$.... 
                 & $\equiv0$ & $-1.1\pm15.6~$ & & $\equiv0$ & $0.1\pm5.3$
                 & & $\equiv0$ & $0.0\pm1.3$\\ 
\hline 
$\chi^2_\up{ring}$........
         & 54.9 & 47.8 & & 20.9 & 15.9 & & 4.1 & 4.1\\
$\chi^2_\up{image}$.....
         & ~0.3 & ~0.1 & & ~0.2 & ~0.2 & & 2.3 & 2.3\\
$\chi^2_\up{lens}$........
         & ~0.3 & ~1.8 & & ~1.3 & ~0.4 & & 0.7 & 0.7\\
$\chi^2_\up{tot}$.........
         & 55.5 & 49.6 & & 22.5 & 16.4 & & 7.1 & 7.1\\
$N_\up{dof}$.........
         & 69 & 63 & & 70 & 64 & & 66 & 60\\
$F$-Test(\%)
         & $-$ & 93.4 & & $-$ & 35.9 & & $-$ & 70.4\\
\enddata
\tablecomments{The best-fit ellipsoidal and nonellipsoidal
models. Only astrometric constraints are used in the fits.
Angles are standard position angles while the lens position and 
the coefficients of the lens potential are calculated in Cartesian coordinates 
in which $x$-direction is toward the West.
The model parameters are the axis ratio $q_s$ and major-axis position angle 
$\theta_s$ of the source, the external shear 
$(\gamma,~\theta_\gamma)$, the shear perturbation $\gamma_g$
produced by a nearby galaxy at ($r_g,~\theta_g$), the position of the lens
$(x_L,~y_L)$, the axis ratio $q_L$ and major-axis position angle 
$\theta_L$ of the lens, the ellipsoidal lens potential 
$\{a_0$, $a_2$, $b_2\}$, and the higher-order deviations $\Delta a_m$ and
$\Delta b_m$ for $m=3,4,5$ from an ellipsoid.}
\label{tab:best}
\end{deluxetable*}

\section{Lens Model}
\label{sec:model}
We use the scale-free potential with arbitrary angular structure
\begin{eqnarray}
{1\over r}\phi(r,\theta)\equiv F(\theta)
&&=a_0+a_2\cos2\theta+b_2\sin2\theta \nonumber \\
&&+\sum_{m=2}^\infty{a_0\over1-4m^2}a_{2m}^q\cos2m(\theta-\theta_L) \nonumber\\
&&+\sum_{m=3}^\infty\left[\Delta a_m\cos m\theta+\Delta b_m\sin m\theta\right],
\end{eqnarray}
as our lens model.
The model has a flat rotation curve, which also seems to be true of the typical
lens galaxy (e.g., \citealt{treu,david2}). This lens model 
has been used extensively
in other studies \citep{zhao,wyn1,wyn2,chris2,wuc,ring,congdon}. 
The model parameters are $a_0$, $a_2$, $b_2$, $\Delta a_m$ and $\Delta b_m$
with $m\geq3$. The coefficients $a_{2m}^q$ for the ellipsoidal part of 
the surface density of the lens galaxy are determined from the
quadrupole moment of the galaxy defined by the model parameters as
\begin{equation}
a_2^q=-{3\over a_0}\left(a_2\cos2\theta_L+b_2\sin2\theta_L\right),
\end{equation}
and the major axis orientation, $\theta_L=0.5\tan^{-1}(b_2/a_2)$.
Note that the $m=1$ terms in this potential are degenerate with
a shift in the source position and can be neglected.
The standard parameters for the deviation 
of the mass density of the lens galaxy
from an ellipsoid are 
$a^\up{B}_m/a_0\equiv(1-m^2)\Delta a_m/a_0$ and 
$b^\up{B}_m/a_0\equiv(1-m^2)\Delta b_m/a_0$ for $m\geq3$ 
(e.g., \citealt{iso1,iso2}). We consider either ellipsoidal models 
($\Delta a_m\equiv\Delta b_m\equiv0$) 
or nonellipsoidal models where $\Delta a_m\neq0$ 
and $\Delta b_m\neq0$ for $m\leq5$. In either type of
model,
we expand the ellipsoid to order $m=5$, beyond which higher-order coefficients 
and deviations are negligible for the relatively round lens galaxies 
(of axis ratio $q_L\simeq1$) we consider here.

We also include two sources of external perturbations. The first is an 
independent external shear due to objects distant from the lens or along the
line-of-sight \citep{chuck1}. 
Thus, we add a term to the potential
\begin{equation}
\phi_\gamma(r,\theta)={1\over2}\gamma r^2\cos2(\theta-\theta_\gamma)\equiv r^2
G_\gamma(\theta),
\end{equation}
with two free parameters for the shear amplitude $\gamma$ and orientation
$\theta_\gamma$. The second source of perturbations are galaxies 
near the main lens that can produce higher-order perturbations beyond an 
external shear (e.g., \citealt{chuck2,chrisnew}).
We approximate these galaxies by the potential,

\begin{eqnarray}
\phi_\up{env}(r,\theta)&&\simeq r\sum_{m=1}^\infty a_m^g\cos m(\theta-\theta_g)
+r^2\sum_{m=1}^\infty b_m^g\cos m(\theta-\theta_g) \nonumber \\
&&\equiv rF_\up{env}(\theta)+r^2G_\up{env}(\theta),
\end{eqnarray}
where the coefficients $a_m^g$ and $b_m^g$ for $m\leq3$ are determined by 
minimizing the difference between
this model and the deflection produced by a
singular isothermal sphere (SIS),
$\phi_\up{SIS}(\bdv{r})=b_g|\bdv{r}-\bdv{r}_g|$, where $b_g$ is the Einstein
radius of the nearby galaxy and $\bdv{r}_g$ is its position vector from the
lens center. The approximation
is accurate to within $0\farcs001$ over an annulus of width $1\arcsec$ 
encompassing the Einstein ring for the closest perturbing galaxy in our sample
(galaxy G22 in HE~0435$-$1225, which is $4\farcs5$ from the lens galaxy).

The total potential is
$\phi(r,\theta)=r\left[F(\theta)+F_\up{env}(\theta)\right]+r^2\left[G_\up{env}
(\theta)+G_\gamma(\theta)\right]$, which may include contributions from several
nearby galaxies. For an ellipsoidal host galaxy with a 
monotonically decreasing surface brightness, the Einstein ring curve defined 
by the peak surface brightness of the Einstein ring as a function of
angle $\theta$ going around the lens galaxy is simply 
\begin{equation}
r(\theta)={\bdv{h}\cdot\mathbf{S}\cdot\bdv{t}+\bdv{u}_\up{o}
\cdot\mathbf{S}\cdot\bdv{t}\over\bdv{t}\cdot\mathbf{S}\cdot\bdv{t}},
\end{equation}
where $\bdv{h}\equiv(F+F_\up{env})\hat{\bdv{e}}_r+(F'+F'_\up{env})
\hat{\bdv{e}}_\theta$, the source plane tangent vector is 
$\bdv{t}=\mathbf{M}^{-1}\cdot\hat{\bdv{e}}_r=(1-2G_\up{env}-2G_\gamma)
\hat{\bdv{e}}_r-(G'_\up{env}+G'_\gamma)\hat{\bdv{e}}_\theta$, 
$\mathbf{M}^{-1}$ is the inverse magnification tensor,
the source center is $\bdv{u}_\up{o}$, 
and $\mathbf{S}$ is the two-dimensional shape tensor of the source
with axis ratio $q_s$ and position angle of its major axis $\theta_s$
(see \citealt{chris2,ring}).

For each lens we fit the image positions, the lens galaxy position,
and the Einstein ring curve
using a simple $\chi^2$ statistic,
\begin{equation}
\chi^2(\bdv{p})=
\sum_i^{N_\up{img}}{|\left[\bdv{u}_\up{o}-\bdv{u}_i(\bdv{p})\right]M_i|^2
\over\sigma_{i,\up{img}}^2}+\sum_i^{N_\up{ring}}{|r(\theta_i)-r(\theta_i;
\bdv{p})|^2\over\sigma_{i,\up{ring}}^2},
\label{eq:chisq}
\end{equation}
where $\sigma_{i,\up{img}}$ and $\sigma_{i,\up{ring}}$ are the uncertainties 
in the $i$-th image position and the $i$-th Einstein ring radius, respectively.
To allow rapid parameter searches, we use this source plane statistic for 
the $\chi^2$ weighted by magnification $M=|\mathbf{M}|$ to approximately 
correct for the difference between deviations on the source and image planes.
Our fiducial ellipsoidal model has 11 free parameters, $\bdv{p}$; 
the source position is $\bdv{u}_\up{o}=(u_x$, $u_y)$, the axis ratio 
$q_s$ and major-axis position angle $\theta_s$ of the source, the ellipsoidal
lens potential $\{a_0$, $a_2$, $b_2\}$, the position of the lens
$(x_L,~y_L)$, and the external shear amplitude and direction 
$(\gamma,~\theta_\gamma)$. In addition, there is one parameter 
$\gamma_g\equiv b_g/2r_g$ for each nearby galaxy with observed position 
$\bdv{r}_g=(r_g,~\theta_g)$ that may be added depending on the environment of 
the lens galaxy. Although flux ratios can provide additional constraints,
they can be changed by dust extinction (e.g., \citealt{dust}),
microlensing (e.g., \citealt{micro}),
and perturbations by substructure (e.g., \citealt{chris3}),
so we will use only astrometric constraints. We used the
Levenberg-Marquardt method to minimize the $\chi^2$ statistic of a given model 
and to estimate the parameter uncertainties, and we checked the convergence
of our solutions by repeating the minimization with a wide range of 
initial parameters and whether we have reached a global minimum by 
using the complementary downhill simplex method (e.g., \citealt{nr}). 

\begin{figure*}[t]
\vspace{-1cm}\centerline{\epsfxsize=7truein\epsffile{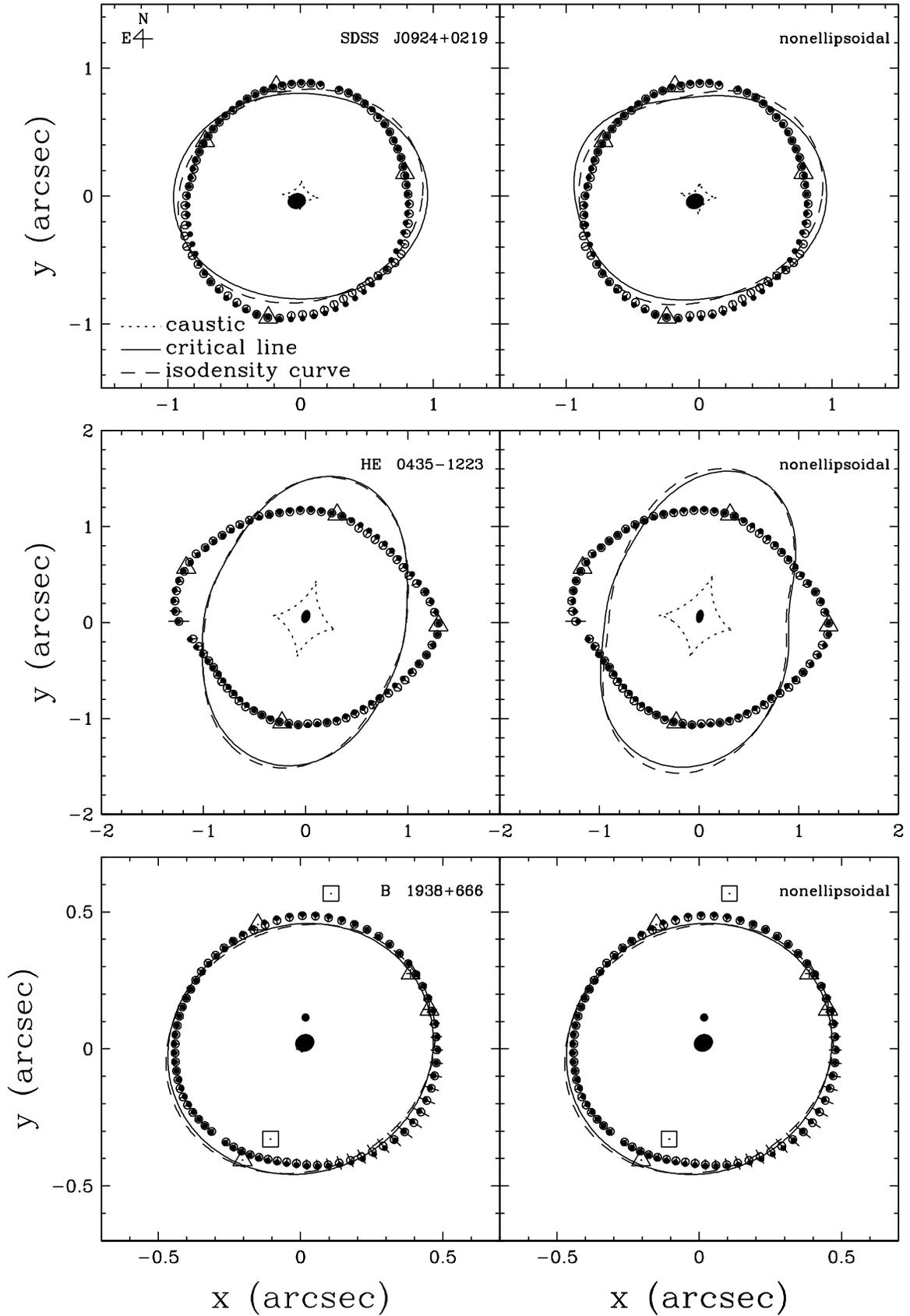}}
\caption{Best-fit ellipsoidal ({\it left panels}) and nonellipsoidal
({\it right panels}) models. Open circles and triangles
represent the observed positions of the Einstein ring and the lensed images.
Positional uncertainties in observations are indicated by the lines passing 
through each symbol. Filled circles show the Einstein ring position predicted 
by best-fit models, while the filled ellipse at the center shows the predicted
position, ellipticity, and position angle of the source. The caustic, the 
critical line, and the isodensity curve of the ellipsoidal model are shown
as dotted, solid, and dashed lines, respectively. 
Open squares in the bottom panels represent the observed
positions of the two additional lensed images, and the small 
filled circle just above
the center shows the predicted position of the additional source for these 
images. Note that the other source position is nearly identical to
the center of the host galaxy ($|\Delta\bdv{u}|\simeq0\farcs023$).}
\label{fig:all}
\end{figure*}

\begin{figure*}[t]
\centerline{\epsfxsize=5truein\epsffile{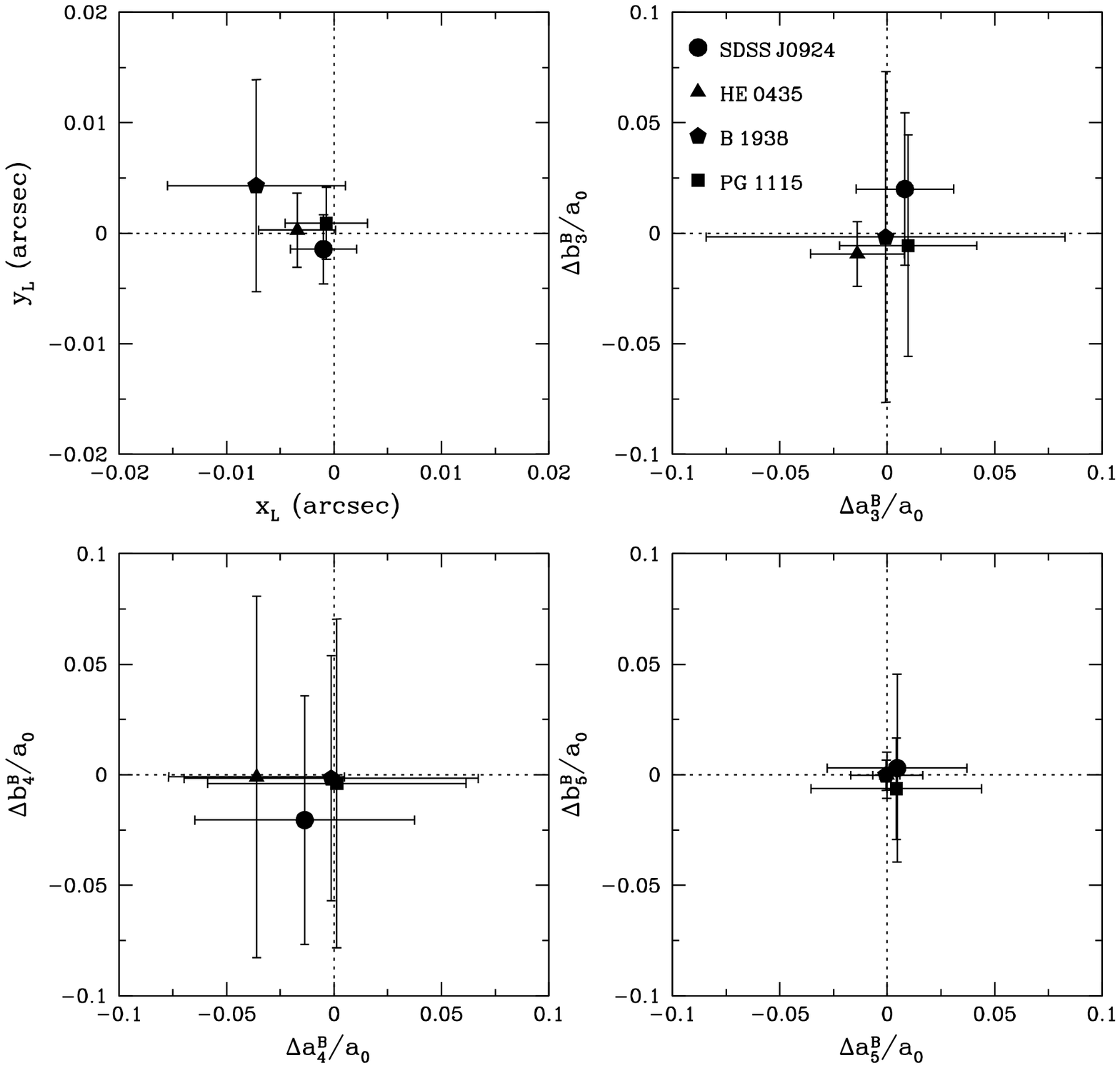}}
\caption{Center of mass offsets ({\it top left}) and the higher-order deviation
coefficients of the best-fit nonellipsoidal models.}
\label{fig:coef}
\end{figure*}

\section{The Structure of Lens Galaxies}
\label{sec:result1}
In this section, we analyze the lens systems to investigate whether the lens
galaxies show any evidence for nonellipsoidal structures in 
their density distribution. As in \citet{ring}, we first fit each lens as an 
ellipsoid centered on the center of brightness of the lens galaxy
and then fit it allowing deviations from ellipsoidal structure or offset
positions. The significance of the changes can be evaluated by using the
$F$-test to estimate the significance of the improvements to the fit 
from adding the additional structural parameters and
by considering the scale of the deviations from ellipsoidal structure.
The best-fit parameters and the $F$-test results for each model and each lens
galaxy are presented in Table~\ref{tab:best}. 
Each lens is modeled with the environmental components (shear, nearby galaxies)
described in \SSS\ref{sec:obs}. We focus on higher order ($m\geq3$) angular
structure in \SSS\ref{sec:ang} and on the match between the center of mass and
the center of light in \SSS\ref{sec:cen}.

\subsection{Angular Structure}
\label{sec:ang}
Our basic result is that SDSS~J0924+0219, HE~0435$-$1223 and B~1938+666 appear
to be ellipsoids, as we had previously found for PG~1115+080 \citep{ring}.
In all four cases, the improvement in the goodness of the fit from adding 
the $m=3-5$ terms to the 
potential is statistically insignificant. The results of the fits are 
summarized in Table~\ref{tab:best}. 

Figure~\ref{fig:all} illustrates the differences in the critical lines and
isodensity contours for the ellipsoidal and nonellipsoidal models of the
three lenses. Figure~\ref{fig:coef} displays the values of the center of 
mass offsets and the higher order
deviations $b_m^\up{B}/a_0$ ($m=3,4,5$)
from an ellipsoid. In Table~\ref{tab:best},
we quantify the difference of the two best-fit ellipsoidal and nonellipsoidal
models by providing the $F$-test probability for the
nonellipsoidal model being consistent with the ellipsoidal model. In none of 
the lenses do the additional structural parameters of the nonellipsoidal model
improve the fits. 

Although we generally ignore the constraints on the flux ratios 
in this paper, we note that the best-fit ellipsoidal models reproduce
the observed flux ratios except for the images with flux anomalies
that are probably due to substructure or stellar microlensing.
For example, the flux anomaly A$_1$/A$_2$ of PG~1115+080 turns out to be due to
the stellar microlensing \citep{micro}. The image D of SDSS~J0924+0219 and
the images A and C of HE~0435$-$1223 also show deviations of predicted
compared to observed fluxes while the global lens models in both cases 
provide a reasonable fit to
the image positions and the flux ratios of the rest of the images.
Since the magnification is more sensitive to the higher order structure in 
the potential, the best-fit nonellipsoidal models using
only astrometric constraints fit the observed flux ratios of the
anomalous images poorly. 
However, even if we impose the flux ratio constraints including the
anomalies, the best-fit parameters change little from those for the ellipsoidal
model, and it is simply impossible to fit the flux ratios while simultaneously
obtaining a good fit to each Einstein ring. 
We consider models based only on the astrometric constraints in the rest of
this paper.

Ellipsoidal models fit the observed image positions and the Einstein ring
of SDSS~J0924+0219 well with $\chi^2_\up{dof}=0.80$. While a slightly better 
fit with $\chi^2_\up{dof}=0.79$ is found using the nonellipsoidal model,
the $F$-test probability that the additional variables were unnecessary is
93.4\%, so the astrometric constraints from the Einstein ring strongly favor
the ellipsoidal model for the lens galaxy of SDSS~J0924+0219 over
nonellipsoidal models. Contrary to \citet{cosmo}, we have no difficulty
finding a model that simultaneously fits the quasar images and the Einstein
ring. The difference probably arises because \citet{cosmo} compared model
parameters from two separate fits to each constraint rather than fitting both
constraints simultaneously.

HE~0435$-$1223 is also well fitted by the ellipsoidal 
model with $\chi^2_\up{dof}=0.32$ (see Fig.~\ref{fig:all}).
In fact, we overfit the data and/or the
uncertainties in the Einstein ring curve are overestimated by 76\%. The 
ellipsoidal model with a nearby SIS and an external shear provides a 
fit consistent
with the recent measurements of the image time delays \citep{chrisnew},
and the parameters of the best-fit nonellipsoidal model are consistent with 
those of the ellipsoidal model. In addition to the $F$-test indicating that 
the additional variables of this nonellipsoidal model are unnecessary,
the best-fit nonellipsoidal model requires a high external shear of
$\gamma=0.065$ compared to 0.035 for
the ellipsoidal model. Considering that we have
taken out the contribution from the nearby galaxy G22, it is 
unphysical to have external perturbations of $\sim$7\%. The high amplitude of 
the external shear in the nonellipsoidal model compensates for the higher order
deviations of the lens potential from an ellipsoid. 

The lens system B~1938+666 exhibits doubly lensed ({\it open squares})
and quadruply ({\it open triangles}) lensed radio images
whose sources are two radio lobes rather than a central core \citep{king}. 
Note that the double radio images
are offset from the Einstein ring.
For this lens, we therefore include two additional source 
positions for the radio images offset from the center of the host galaxy 
producing the Einstein ring. The nearly circular Einstein ring rules out
any significant higher order deviations in nonellipsoidal models, and the 
best-fit nonellipsoidal model is consistent with the ellipsoidal model
(see Fig.~\ref{fig:coef}).

\subsection{Center of Dark Matter Halos}
\label{sec:cen}
In the previous subsection we constrained the position of the lens galaxy to
agree with the position of the luminous galaxy seen in the $HST$ images. 
This is the standard assumption of almost all lens models. In principle,
the center of mass could differ from the center of light if the halo were
genuinely offset from the stars, if the halo were significantly lopsided, or
if there were any sufficiently massive dark substructures inside the Einstein
ring.
To test these possibilities, we drop the constraints on the lens position 
and repeat the optimization of the ellipsoidal models to estimate the offset
$(x_L,~y_L)$ of the center of mass from the lens galaxy.
Physically it seems unlikely that
the galaxy light can be significantly offset from the dark matter halo
on galaxy scales, and this is borne out by these results.
We find that the center of mass can be offset from the center of
light by only $\Delta_\up{cm}=0\farcs001\pm0\farcs003$, 
$0\farcs003\pm0\farcs004$, $0\farcs008\pm0\farcs009$, and
$0\farcs001\pm0\farcs002$ for SDSS~J0924+0219, 
HE~0435$-$1223, B~1938+666, and PG~1115+080, respectively. These correspond
to physical offsets of $\Delta_\up{cm}=5.4\pm16.3$, 17.4$\pm23.1$, 
61.9$\pm69.6$, and 4.6$\pm9.1$~pc. The nonellipsoidal models only increase
the uncertainties in the offsets by roughly a factor of 1.5
and provide upper limits consistent with
the values of the offsets. For a typical dark matter 
fraction inside the Einstein ring of $\epsilon\simeq30\%$ \citep{koop,david},
the offset in the center of mass of the dark matter halo
is related to the offset in the center of mass
by $\Delta_\up{dm}\simeq\Delta_\up{cm}/\epsilon$, so the limits on the offset
of the dark matter
halo are somewhat weaker. Note that this co-location of the stars and the
halo may not hold in clusters. For example, models of the cluster lens 
SDSS~J1004+4112 require an offset of $\gtrsim5.7~\kpc$
between the cD galaxy and the center of the cluster potential
\citep{masa}, and there is some evidence that cD galaxies are not at rest with
respect to the rest frame defined by the other cluster galaxies (e.g.,
\citealt{frank}). 

The offsets also test the lopsidedness of dark matter
halos and the existence of massive dark substructures. 
The luminous galaxies show no evidence for lopsidedness,
because they are well modeled by ellipsoidal de Vaucouleur profiles.
Since dark matter
halos are also supported by random particle motions, they should
also have difficulty supporting a lopsided density structure, and this is 
borne
out by the tight limits on $\Delta_\up{dm}$. Finally, the lack of an offset
also sets a limit on the existence of massive dark substructures. Given a 
typical subhalo mass function, most massive substructures dominate the
total mass contained in substructure \citep{and,ben,taylor,gao,zentner05}. 
If we add a single object whose mass is fraction
$f_\up{sat}$ of the mass inside the Einstein ring at radius $r_\up{sat}$ for
the lens galaxy, it would shift the center
of mass by $\Delta_\up{cm}=f_\up{sat}r_\up{sat}$. Thus, we have a limit
that $f_\up{sat}\lesssim\Delta_\up{cm}/\langle r_\up{sat}^2\rangle^{1/2}$ 
where 
$\langle r_\up{sat}^2\rangle^{1/2}$ is the rms distance of the 
satellites inside the Einstein ring from the lens center. Crudely, 
$\langle r_\up{sat}^2\rangle^{1/2}\simeq b$, the Einstein 
radius, so $f_\up{sat}\lesssim1\%$. Adding additional satellites only weakly
affects the limits, increasing
the limit on $f_\up{sat}$ by $\sqrt{N}$ for $N$ equal mass satellites.
This strong sensitivity to massive
substructures is presumably one of the reasons that 
it is almost impossible to model lenses like MG~J0414+0534 with visible
satellites using only a central potential (e.g., \citealt{ros,mg0414}).

\begin{figure}[b]
\centerline{\epsfxsize=3.5truein\epsffile{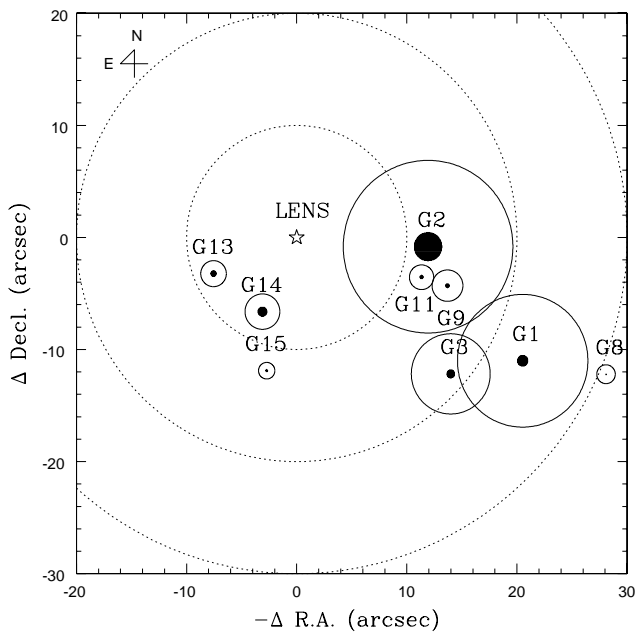}}
\caption{The environment of PG~1115+080. The nearby galaxies are 
represented by the labeled circles, where the sizes of the circles are 
proportional to the perturbation
produced by each galaxy. Solid circles are for the
external shear, and filled circles are for the 
higher-order perturbations. The dotted
concentric circles centered on the main lens galaxy
are reference circles with radii of 10\farcs0, 20\farcs0, and 30\farcs0.}
\label{fig:env}
\end{figure}

\section{Group vs. Galaxy Halos in PG~1115+080}
\label{sec:result2}
In the halo model that describes the relation of galaxies and dark 
matter,  a luminous central galaxy 
is located at the center of dark matter halo and the satellite
galaxies are distributed around the central galaxy
(e.g., \citealt{dhw,andrey}).
In \SSS\ref{sec:cen}, we found that visible lens galaxies must be 
centered on their halos, and that they generally lack massive substructures.
If, however, we consider a lens consisting of a galaxy in a group, like 
PG~1115+080, then we can
explore the relationship between central and satellite galaxies.
This is potentially important 
because the time delay measurements for PG~1115+080 \citep{paul}
require a main
lens galaxy with almost no dark matter at the concordance value
of the Hubble constant $H_0=72~\kms\up{Mpc}^{-1}$ \citep{chris0}, in marked
contrast to the typical lens galaxies, which seem to have a significant
dark halo (e.g., \citealt{treu,david2}). This case
suggests that the structures of early-type galaxies are heterogeneous.
One possible interpretation of this heterogeneity is that the primary lens of 
PG~1115+080 is a satellite lens galaxy
 of the group that has been stripped of most
of its dark matter. In the context of the halo model, 
this implies that one of the 
other visible galaxies around PG~1115+080 must be the central galaxy of the 
group, and it should show a significantly higher mass-to-light ratio than
the other galaxies, all of which are 
satellites. In this section, we attempt to interpret PG~1115+080 within this
theoretical picture.

\subsection{The Environment of PG~1115+080}
Figure~\ref{fig:env} shows the environment of PG~1115+080.
Earlier models with an independent external shear or an SIS
external perturber consistently indicate that the lens has a $\sim$10\% shear
perturbation associated with its parent group,
with weak evidence that the group
halo should be located near the luminosity weighted position of the nearby 
galaxies
\citep{paul,keeton,courbin,impey,zhao,chris3,ring}. The 
galaxies G1, G2 and G3 are comparable in luminosity to the main
lens galaxy, and
the less luminous galaxy G14 is the closest. To estimate the 
perturbations produced
by each galaxy, we use the $I$-band luminosities $L$ of the nearby galaxies
in Table~\ref{tab:env} \citep{impey} and the Faber-Jackson relation 
to estimate their stellar 
velocity dispersions $\sigma_\star\propto L^{1/4}$ and
the Einstein radii $b\propto\sigma_\star^2$ 
($L\propto\sigma_\star^4\propto b^2$).
Scaling all the galaxies based on the luminosity and Einstein radius of the
main lens galaxy, we then compute the relative
amplitudes of the external shear $\propto b/r_g$ 
and the nonlinear perturbation $\propto b/r_g^2$ produced by each galaxy 
given its distance $r_g$ from the lens (see Table~\ref{tab:env}).
As illustrated in Figure~\ref{fig:env}, the
luminous galaxies G1, G2, and G3 dominate the perturbations while the
nearby, less luminous galaxies G13 and G14 make a non-negligible contribution. 

\begin{deluxetable}{lrrcc}
\tablewidth{0pt}
\tablecaption{Galaxy Environment of PG~1115+080}
\tablehead{\colhead{Object} & \colhead{$I$(mag)} & \colhead{$r_g$(arcsec)} & 
\colhead{$\theta_g$(degrees)} & \colhead{$\gamma_g$} }
\startdata
Lens..... &  $18.55\pm0.50$ & $\equiv0$ & $-$      & $-$   \\ 
G1........ & $17.85\pm0.01$ & $23.285$  & $241.80$ & 0.030 \\ 
G2........ & $18.73\pm0.04$ & $11.957$  & $266.05$ & 0.039 \\ 
G3........ & $19.44\pm0.02$ & $18.543$  & $228.97$ & 0.018 \\ 
G8........ & $21.52\pm0.08$ & $30.624$  & $246.52$ & 0.004 \\ 
G9........ & $19.44\pm0.02$ & $14.352$  & $252.57$ & 0.007 \\ 
G11...... &  $22.98\pm0.34$ & $11.881$  & $252.62$ & 0.006 \\ 
G13...... &  $23.70\pm0.10$ & $8.214$   & $113.16$ & 0.006 \\ 
G14...... &  $23.30\pm0.10$ & $7.313$   & $154.82$ & 0.008 \\ 
G15...... &  $23.80\pm0.10$ & $12.198$  & $167.09$ & 0.004 \\ 
\enddata
\tablecomments{Galaxy positions are relative to the lens.
We use the $I$-band luminosities and the Faber-Jackson relation to estimate
the relative amplitudes of the external shear $\gamma_g$ produced by each
galaxy. The nonlinear perturbations are smaller than $\gamma_g$ by $r_g$.
The positions and luminosities are from \citet{impey}.}
\label{tab:env}
\end{deluxetable}

\begin{figure}[b]
\centerline{\epsfxsize=3.5truein\epsffile{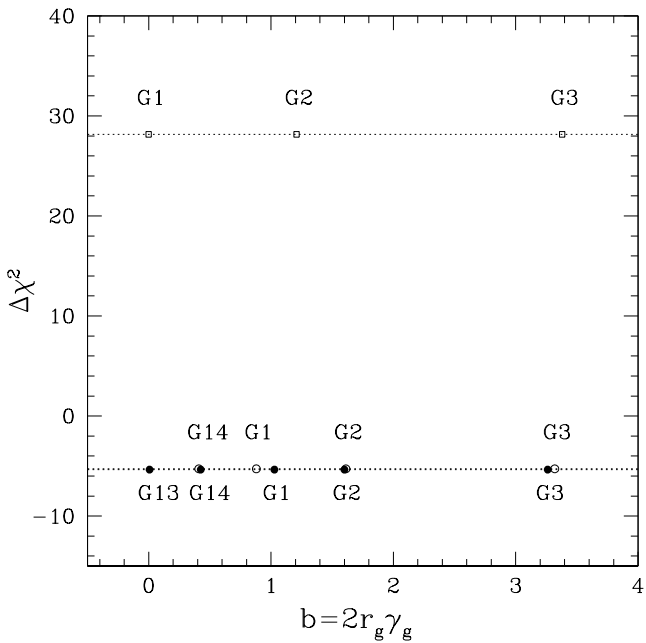}}
\caption{Best-fit models for the environment in PG~1115+080. The 
goodness-of-fit relative to the best-fit model with a group halo
unassociated with the visible galaxies
is shown with the
Einstein ring radii of galaxies in each model.
The environment is approximated as three ({\it open squares}), four 
({\it open circles}), and five nearby galaxies ({\it filled circles}) from top
to bottom. Note that the models with four and five nearby galaxies are 
degenerate.}
\label{fig:4sis}
\end{figure}

We first test whether putting mass only at the positions of the visible 
galaxies can adequately fit the data, initially with no extra external shear.
The galaxy positions are measured to such high accuracy that we fix the 
observed position $\bdv{r}_g=(r_g,~\theta_g)$ for each galaxy,
while we vary their Einstein radii
$b_i$ through the free parameter $\gamma_{g,i}$.
We started with a model using just the three bright galaxies G1, G2, and G3. 
Figure~\ref{fig:4sis} shows the best-fit models.
The three-galaxy environment fits the data quite poorly compared to 
standard models with a group halo centered near the center of light of the 
galaxies ($\Delta\chi^2\simeq28$). Only if we allow a large, independent
external shear ($\gamma\gtrsim0.07$) can we obtain a good fit.
However, when we include the next 
leading perturber, G14, the goodness of fit is somewhat
better ($\Delta\chi^2\simeq-5$) than for
the fit with a halo unassociated with the visible 
galaxies. Note however that this new model has one more degree of freedom 
compared to the standard
model, so the improvement in the fit is physically interesting but not
statistically significant.
Including the fifth most important galaxy, G13, hardly changes
the goodness-of-fit, and the Einstein radii of the galaxies change little
(although there are mass degeneracies between G1 and G2/G3).
If we add an independent external shear to the four or five galaxy models,
the estimated 
shear amplitude is small ($\gamma\simeq0.01$), and the goodness-of-fit 
and the Einstein radii of the galaxies remain unchanged.
The addition of any more nearby galaxies simply contributes to
the external shear $\gamma$
because they produce negligible higher-order perturbations.
Therefore, we adopt
the four-galaxy model of G1, G2, G3, and G14 with an external shear of
amplitude $\gamma<0.01$ as our
fiducial model for the environment in PG~1115+080. Next we address whether
the masses of these galaxies show any correlations that test the structures
of their halos.

\subsection{Halo Structure of the Galaxy Group of PG~1115+080}
Our parameter for each galaxy, $\gamma_g\propto M(<r_g)/r_g^2$, is proportional
to the projected mass $M(<r_g)$ of each galaxy out to the distance of the 
galaxies $r_g$
from the lens. Based on the halo model, our working hypothesis 
is that one of these four galaxies is the central galaxy of the halo and
should have different halo properties than the 
other satellite galaxies. 

For any fixed assumption about the halo structure and assuming the galaxies
have similar mass-to-light ratios, we would expect to see a correlation between
these mass estimates $b_i$ and the galaxy luminosities $L_i$. 
If there is a central galaxy
with a very different halo mass than the satellite galaxies, then it should be
identifiable as the galaxy least in agreement with the correlation of the 
other galaxies. We first consider two limiting cases in which the halos
are either all truncated on a scale less than the distance to the lens or where
all halos conspire to have flat rotation curves out to the distance to the 
lens. The Einstein radius is $b_i^\up{est}\propto L_i^{1/2}$ in both cases 
while the shear amplitude produced by each galaxy is 
$\gamma_i=b_i^2/r_{g,i}^2$ in the former case and $\gamma_i=b_i/2r_{g,i}$ 
in the latter case.
We include the correlation by adding a term to the fit statistic of the form
\begin{equation}
\Delta\chi^2_{M-L}=2(N_\up{gal}-1)\ln\sigma+
\sum_{i=1}^{N_\up{gal}}{\left[\ln b_i-\ln b_i^\up{est}(L_i)\right]^2 \over 
\sigma^2},
\label{eq:p1}
\end{equation}
where $\sigma$ is the logarithmic scatter in the correlation and 
$N_\up{gal}=4$. 

\begin{figure}[t]
\centerline{\epsfxsize=3.5truein\epsffile{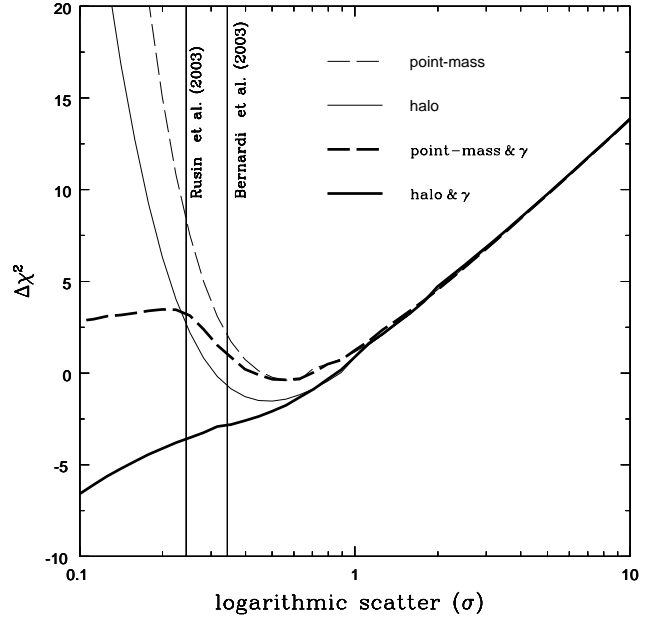}}
\caption{Mass models for the galaxy environment. We assume a Faber-Jackson 
relation $L\propto b^2$ with a natural logarithmic scatter $\sigma$. We show 
the goodness-of-fit as a function of the scatter for
each mass model relative to the best-fit model of the four 
nearby galaxies with arbitrary mass. 
Solid lines show the results for galaxies with extended halos, and dashed lines
show the results for point-mass halos. The light curves allow no additional
external shear while the thick lines allow a small external shear.
Since the mass scales of the galaxies are not in 
perfect agreement with the correlation in the mass models without an external
shear, the difference in $\chi^2$ becomes substantial as a tight correlation
is imposed ($\sigma\ll1$). The mass models with an external shear can perfectly
agree with the correlation, and the $\Delta\chi^2$ is controlled by the assumed
prior on the scatter. Two vertical lines show the logarithmic scatters found 
in the SDSS for nearby elliptical galaxies \citep{sigma} and for ensembles of
lens galaxies \citep{rusin}.}
\label{fig:rel}
\end{figure}

Figure~\ref{fig:rel} compares the two mass models with and without an external
shear $\gamma$. When we include the shear, we use a prior constraint of
$\gamma=0.00\pm0.01$ (based on the previous four galaxy model).
The model assuming extended halos
fits the data slightly better ($\Delta\chi^2\simeq2$)
than the point-mass model regardless of the assumed scatter or the addition of
the external shear. The logarithmic scatter of the correlation is 
$\sigma\simeq0.5$ for the best-fit model, and will give $\sim65\%$ fractional 
differences. Note that there is little evidence of a useful correlation,
but the scatter is not significantly different from 
the Faber-Jackson relation between the luminosity
$L$ and the central velocity dispersion
$\sigma_\star$ observed for early-type galaxies in the SDSS 
($\sigma\simeq0.34$ in our units, \citealt{sigma}) or that measured from
ensembles of lens galaxies ($\sigma\simeq0.24$, \citealt{rusin}).
When an external shear is allowed, 
it turns out that there is degeneracy between the external
shear and the mass models for the four galaxies such that all the
galaxies can perfectly agree with the assumed correlation, and the results
are controlled by our prior on the scatter.
The extended halo model requires $\gamma\simeq2\%$ to fit the data with all
galaxies on the correlation, while the point-mass model needs 
$\gamma\simeq4\%$. 

As our next experiment, we allow one of the
four galaxies to have an arbitrary mass while imposing the correlation on
the rest of the three galaxies ($N_\up{gal}=3$).
Unfortunately, no single
galaxy stands out, and the best-fit models without an external shear 
again require large intrinsic scatters 
$\sigma\simeq0.5$. When an external shear is allowed in the
model, the external shear can always compensate for
the shear produced by the galaxies due to
the degeneracy between the external shear and the mass models,
and no models are significantly better than the models 
with the four galaxies in the correlation. 

\section{Summary}
\label{sec:sum}
In this paper we explored the structure of halos using
four gravitational lenses with well-defined
Einstein ring images of the quasar host galaxies. Adopting a
scale-free potential with arbitrary angular structure, we 
considered both ellipsoidal and nonellipsoidal gravitational potentials,
searching for any deviation from an ellipsoidal potential.
Based on the astrometric constraints, none of the
additional free parameters represents an improvement over
the best-fit ellipsoidal models,
and the best-fit nonellipsoidal models are consistent with their ellipsoidal
counterparts. 
Furthermore, imposing the observed flux ratios as constraints does not alter
the results. Anomalous flux ratios can not be
eliminated by changing the central potential.
The angular structures of all four lens systems are consistent with 
ellipsoidal models.

We also investigated the offset of the center of mass of the lenses
from the measured center of light by dropping the constraints that the lens
is centered on the visible galaxy. We find that the offsets 
are required to be $\Delta_\up{cm}\lesssim10$~pc. 
This roughly implies that the dark matter halos can be offset
by at most $\Delta_\up{dm}\simeq\Delta_\up{cm}/\epsilon\lesssim30$~pc, 
where $\epsilon\simeq30\%$ is
a typical fraction of dark matter inside the Einstein ring.
This may be markedly different from clusters
in which the cD galaxy can be significantly offset from the center of the
cluster potential. Dark substructures of mass fraction
$f_\up{sat}$ could also shift the center of mass if there are few of them,
and this allows us to set a limit of
$f_\up{sat}\lesssim\sqrt{N}\%$ for $N$ equal mass substructures inside the 
Einstein ring. This limit is not tight enough to represent a conflict
with either theoretical \citep{taylor,gao,zentner05}
or observational \citep{neal1} estimates of the
substructure fraction.

Finally, we explored the environment of PG~1115+080 to study
the halo structure of its parent group under the assumption 
that the individual dark matter 
halos are centered on the nearby luminous galaxies.
We focused on the four galaxies G1, G2, G3, and G14 producing the largest
perturbations, since the other nearby galaxies 
produce negligible higher-order perturbations and have effects
degenerate with an independent external shear. 
Placing the mass solely on these galaxies can fit the data, but we find 
only weak evidence for any systematic correlation between the mass and the
luminosity. Models assuming the galaxies have extended halos on the scale of
their separations are slightly better than those where they do not, but the
scatter in the correlation between the Einstein radius and $I$-band 
luminosity must be $\sim65\%$. 
This is comparable to that implied by Faber-Jackson relation observed for
nearby galaxies or for lens galaxies. We also found
no evidence for a marked difference between one of the galaxies (the 
``group halo central galaxy'') and the others (satellite galaxies). The rapid
convergence of the perturbations from nearby galaxies to a single external
shear probably means that lenses like PG~1115+080 provide an insufficient
number of constraints to test the hypothesis of the halo model in detail.

\acknowledgments
This research has been supported 
by grants HST-GO-7495, 9375, and 9744 from the Space Telescope
Science Institute, which is operated by the Association of
Universities for Research in Astronomy, Inc., under NASA
contract NAS 5-26555.

\end{document}